\documentclass[12pt]{article}

\usepackage{graphicx}
\usepackage{natbib}
\def\mnras{MNRAS}
\def\araa{ARA\&A}
\def\prl{Phys.~Rev.~Lett.}
\def\apj{ApJ}
\def\apjs{ApJS}
\def\prd{Phys.~Rev.~D}
\def\aap{A\&A}
 
\begin{document}

\def\stackunder#1#2{\mathrel{\mathop{#2}\limits_{#1}}}

\title{$C^\infty$ perturbations of FRW models with a cosmological constant}
\author{Zolt\'an Perj\'es$^{1}$, M\'aty\'as Vas\'uth$^{1}$, Viktor Czinner$^{1}$
\\ and Daniel Eriksson$^{2}$ \\ 
{\small ${}^{1}$KFKI Research Institute for Particle and Nuclear Physics,}\\ 
{\small Budapest 114, P.O.Box 49, H-1525 Hungary}\\
{\small ${}^{2}$ Department of Physics, Ume\aa\ University, SE-90187 Ume\aa, Sweden}}

\date{\today}

\maketitle

\begin{abstract}
\noindent 
Spatially homogeneous and isotropic cosmological models, with a
perfect fluid matter source and non-vanishing cosmological constant, are studied. The
equations governing linear perturbations of the space-time and the variation
of energy density are given. The complete solution of the problem is obtained for $C^{\infty}$
perturbations, using a comoving time. The Sachs-Wolfe fluctuations of the temperature of the
cosmic background radiation are obtained for the relatively growing density perturbations. 
It is found that the observable celestial microwave fluctuation pattern underwent a reversal
approximately two billion years ago. What is observed today is a negative image of the last 
scattering surface with an attenuation of the fluctuations, due to the presence of the 
cosmological constant.
\vskip 12pt
\noindent
Keywords: {\it cosmic microwave background -- large scale structure of the universe}

\end{abstract}

\section{Introduction}

Recent observations by the {\it High-z Supernova Search Team} \citep{Tonry} corroborate 
the data on an accelerating expansion of the Universe. Three independent lines 
of evidence [those from the {\it Wilkinson Microwave Anisotropy Probe} ({\it WMAP}) 
measurements \citep{WMAP} of the cosmic 
microwave background  radiation {\it (CMBR)}, the {\it Sloan Digital Sky Survey} ({\it SDSS}) \citep{Tegmark} and the 
 observations of type-Ia supernova spectra] converge on the value 
 $\Omega_{\Lambda}=0.70\pm 0.04$ of dark 
energy. All data are consistent, within a 20\% error bar, 
with a time-independent dark energy distribution, as is the case with the cosmological 
constant $\Lambda$ and flat space ($k=0$). These developments attracted attention to 
cosmological models in the presence of a $\Lambda$ term in Einstein's gravitational equations. 
Many excellent review papers are
available \citep  {Carroll,Peebles,Carroll2,Silk} on the implications 
of a cosmological constant in models of the Universe. While the cosmological constant has little 
effect on the large-scale structure and  dynamics of the Universe, the microwave fluctuations 
will be significantly different in $\Lambda \neq0$ models \citep{Silk}.
The scalar perturbations of $\Lambda \neq0$ models with multicomponent fluids (i.e., cold or hot 
 dark matter, photons and massless neutrino) have been computed by \citet{Bond},
\citet{Fukugita}, \citet{Holtzman}, \citet{Vittorio}, \citet{HS} and \citet{Stompor} in the linear 
approximation. These numerical results have been compared by \citet{Stompor} and found to be in 
reasonable agreement. \citet{HS} claim that for 
$\Lambda\neq0$, one must use a numerical approach. In our work, however, we present
an analytic treatment of the collisionless dust which represents the late-time evolution.
 
 A complete solution describing the effect of the cosmological constant on 
perturbation dynamics is lacking in the literature, this 
including the rotational and gravitational wave perturbations. 
The temperature fluctuations of the CMBR have been computed
in numerical schemes \citep{Bond,Fukugita,Holtzman,Vittorio,Stompor}
and \citet{Multamaki} investigate
the integrated contributions. Known as the Sachs-Wolfe effect, this is the 
variation in the redshift of a photon travelling freely in the universe rippled with
perturbations. Initial work, both on the zero-pressure model of
the current state of the Universe, and on the radiation-dominated era, was done 
by \citet{SW}. The pressure-free universe in the presence of a cosmological
constant has been considered by \citet{Heath} and \citet{Lahav}. They 
used the redshift parameter to investigate the behaviour of the density contrast. A more 
recent work by \citet{Vale} uses the formalism of \citet{Padmanabhan} to compute linear perturbations
of a dust-filled Friedmann-Robertson-Walker (FRW) universe in the presence of a cosmological constant. They find
that the presence of $\Lambda$ inhibits the growth of the fluctuations in time.
[A different approach to the perturbation problem of a different (de Sitter) model, 
using the gauge-independent
formulation, has been pursued by \citet{Barrow}, for the investigation
of stability under perturbations.]

Neither of these earlier works reaches a complete perturbative picture. Although the 
Sachs-Wolfe effect is computed by numerical methods, there is no analytic treatment so far 
available for $\Lambda\neq0$. Our aim here, therefore, is to consider, in an analytic framework,
the Sachs-Wolfe effect and its contributions to the fluctuations of the CMBR in the presence of a
cosmological constant. We follow the treatment of the perturbations due to \citet{White}, 
thereby relaxing the momentum conditions of the original Sachs-Wolfe work. Both works yield 
all $C^\infty$ perturbations although this was only shown by \citet{White}. 
Correspondingly, in the present paper we also obtain all $C^\infty$ perturbations.
In Sec. 2, we enlist the perturbed field equations using the conformal time $\eta$.
While retaining the form of the metric functions, we change to the comoving time $t$ in the 
field equations so obtained. The reason for this is that the unperturbed radius $a$ has a 
simple analytic dependence on $t$.

  The general solution of the linear field equations in a dust-filled universe is obtained in Sec. 3.
We find that all first-order fields, except the wave solutions, can be expressed in terms of only two 
complex incomplete elliptic integrals $E$ and $F$. This is achieved by using the mirror 
symmetries of the elliptic integrals.

  In Sec. 4, we obtain the Sachs-Wolfe effect and its integral contribution which vanishes in
the $\Lambda\rightarrow 0$ limit. An unexpected feature of the transfer function is that it
changes sign and reverses the pattern of the temperature fluctuation on the celestial sphere
at late times. In Sec. 5, we investigate the physical interpretation of our solution. Taking 
the current experimental values of the cosmological parameters, we obtain the comoving time elapsed
between the emission at the surface of last scattering and reception of
the photon. We find that the Sachs-Wolfe contribution to the CMB fluctuations in the presence 
of the cosmological constant is damped by a factor of $3$ today,
because of the proximity, on the cosmological time scale, of the moment of 
reversal of the microwave temperature fluctuations and the reception. This feature of the
SW effect has remained concealed in earlier works, because of the numerical methods involved.

\section{Linear perturbations}

   In this section we present the equations of a perturbed spatially 
flat ($k=0$) FRW cosmology in the linear approximation. 
The metric is that of a perturbed FRW model in the conformal form 
\begin{eqnarray}  \label{FRW}
g_{ab} = a^2(\eta )(\eta_{ab}+h_{ab}) \ ,
\end{eqnarray}
where $\eta_{ab}={\rm diag}(1,-1,-1,-1)$, $h_{ab}$ is the metric
perturbation and the scale function $a=a(\eta )$ is determined by the field
equations. Roman indices run from 0 to 3 and Greek indices from 1 to 3. Except 
when we compute numerical values, we employ units such that the speed of light is $c=1$ 
and the gravitational constant $G=1/8\pi$.
We use the Minkowski metric $\eta_{ab}$ and its inverse to raise and lower indices 
of $h_{ab}$ and of other small quantities.

The matter source is assumed to be a perfect fluid, 
\begin{eqnarray}  \label{enmom}
T^{ab} = (\tilde{\rho} + \tilde{p})u^au^b - \tilde{p}g^{ab}
\end{eqnarray}
with the four velocity $u^a$ of the fluid normalized by $u^iu_i=1$. The
cosmological constant $\Lambda$ of the Einstein equations is introduced in
the energy-momentum tensor. The effective energy density $\tilde{\rho}$ and
pressure $\tilde{p}$ are related to the corresponding fluid quantities as 
\begin{eqnarray}\label{rhop}
\tilde{\rho} = \rho + \Lambda\ ,\quad \tilde{p} = p - \Lambda \ . 
\end{eqnarray}
In the perturbed space-time we choose comoving coordinates \citep{SW}. Using
this gauge, the four velocity of the fluid and the metric perturbation $%
h_{00}$ are 
\begin{equation}  \label{coordcond}
u^a=\frac{\delta^a_{\ 0}}{a}\ , \quad h_{00}=0 \ .
\end{equation}

The coordinate transformations $x^a \rightarrow x^a-\xi^a$, which preserve
the coordinate condition, have the following properties 
\begin{eqnarray}\label{cootr}
\xi^r u^a_{\ ,r}-u^r \xi^a_{\ ,r} = 0 \ \ \Rightarrow \ \ \xi^0 = \frac{%
b(x^\alpha)}{a} \ , \quad \xi^\alpha = c^\alpha(x^\beta) \ ,
\end{eqnarray}
where the arbitrary functions $b$ and $c^\alpha$ are independent of the
conformal time $\eta$.

The conservation of energy-momentum 
\begin{equation}
T^{ab}=(\rho+\delta\rho+p+\delta
p)u^au^b-(p+\delta p-\Lambda)g^{ab}
\end{equation} 
implies  the unperturbed relations
\begin{eqnarray}  \label{aeq}
3(\rho+p)a^\prime + a\rho^{\prime} = 0\ , \quad p_{,\alpha} = 0 \ 
\end{eqnarray}
and the perturbative equations
\begin{eqnarray}  \label{cons}
&&\delta\rho^{\prime}+3\frac{a^\prime}{a}(\delta\rho+\delta p) + (\rho+p)\frac{%
h^{\prime}}{2}=0 \ , \\  \label{cons1}&&\left( h^{\alpha 0\ \prime}+\frac{a^\prime}{a}%
h^{\alpha 0}\right)(\rho+p) +h^{\alpha 0}p^{\prime}-\delta p^{\
\alpha}_{,}=0 \ 
\end{eqnarray}
where $h=\eta^{rs}h_{rs}$.
The prime denotes derivative with respect to the conformal time $\eta$.

 Einstein equations $G_{ab} = T_{ab}$ to leading order give 
\begin{eqnarray}  \label{Field0}
3\frac{a^{\prime 2}}{a^4} = \rho +\Lambda \ , \quad 2\frac{%
a^{\prime\prime}}{a^3}-\frac{a^{\prime 2}}{a^4} = - p +\Lambda \ .
\end{eqnarray}
The perturbed field equations for a perfect fluid have been obtained by \citet{White} . They
remain valid in the presence of the cosmological constant provided the pressure and density are
replaced by their tilded values according to (\ref{rhop}). The scale factor $a$ is now a 
solution of Eqs. (\ref{Field0}).  
Combining the $00$
component and the trace of the spatial part of the Einstein equations we
have 
\begin{eqnarray}  \label{Eq00}
&& h^{\prime\prime} = - 2\frac{a^\prime}{a}h^\prime + 2h^{0\mu}_{\ \ ,
\mu}{}^\prime + 4\frac{a^\prime}{a}h^{0\mu}_{\ \ , \mu}\\\nonumber 
&&\qquad\quad+\frac{1}{2}S^{\mu\nu}_{\ \ , \mu\nu} + \frac{1}{3}\nabla^2h 
- 3a^2\delta p \ , \\&& S^{\mu\nu}_{\ \ ,\mu\nu} + \frac{2}{3}\nabla^2h + 
2\frac{a^\prime}{a}\left(2h^{0\mu}_{\ \ , \mu} - h^\prime \right) + 2a^2\delta\rho = 0 \ .
\label{Traceeq}
\end{eqnarray}
In these equations $S_{\alpha\beta}=h_{\alpha\beta}-\eta_{\alpha%
\beta}h/3$ denote the trace-free part of the perturbations and 
$\nabla^2 f=-\eta^{\mu\nu}f_{,\mu\nu}$, where $\nabla^2\equiv \Delta$
is the standard Laplacian,
for an arbitrary smooth function $f$. 
The remaining components of the linearized Einstein equations are 
\begin{eqnarray}  \label{Eq0al}
&&\nabla^2h^{0\alpha}-\frac{2}{3}h_{,}^{\ \alpha\prime} 
+S^{\alpha\mu}_{\ \ \ ,\mu}{}^\prime \\\nonumber 
&&+ h^{0\mu}_{\ \ , \mu}{}^\alpha 
+4\left(\frac{a^{\prime\prime}}{a} - 2\frac{a^{\prime 2}}{a^2}%
\right)h^{0\alpha} = 0   
\end{eqnarray}
\begin{eqnarray}
&&S_{\ \beta}^{\alpha\ \prime\prime}\! + 2\frac{a^\prime}{a}S_{\
\beta}^{\alpha\ \prime} \! - \! \nabla^2S_{\ \beta}^{\alpha} = S_{\ \ \
,\beta\mu}^{\alpha\mu}\! + S_{\beta\mu ,}^{\ \ \ \alpha\mu} - \frac{2}{3}%
\delta^{\alpha}_{\ \beta}S_{\ \ ,\mu\nu}^{\mu\nu} \nonumber \\
&& +h_{\ \ ,\beta}^{\alpha 0\
\ \prime}+h_{\beta 0,}^{\ \ \ \alpha\prime} -\frac{1}{3}h_{,\beta}^{\ \
\alpha} +2\frac{a^\prime}{a}(h_{\ \ ,\beta}^{\alpha 0}+h_{\beta 0,}^{\ \ \
\alpha})  \label{Eqalbe} \\
&&  -\frac{1}{3}\delta^{\alpha}_{\ \beta}\left[ \frac{1}{3}\nabla^2 h
+2h_{\ \ ,\mu}^{0\mu\ \ \prime}+4\frac{a^\prime}{a}h_{\ \ ,\mu}^{0\mu} %
\right] \ .  \nonumber
\end{eqnarray}

\section{Integration of the field equations}

In this section we solve the field equations in the presence of pressureless 
matter, $p=\delta p=0$ for perturbations in the class of $C^{\infty}$ functions. 
We then repeatedly use the following 

\noindent{\bf Lemma }({\em Brelot})

{\em If $g$ is any $C^\infty$ function on $E^3$, then there exists a $C^\infty$ 
function $f$ on $E^3$ such that
$\nabla^2f=g$ .}

\noindent(A modern proof was provided by \citet{Friedman}). 
We introduce the comoving time coordinate $t$ by the relation $ad\eta=dt$.
The solution of Eqs.(\ref{Field0}) for matter \citep{Stephani} is 
\begin{eqnarray}  \label{backgr}
&&p=0\ ,\quad \rho a^3={\cal C}_M\ ,\quad a=a_0 {\rm sinh}^{2/3}(Ct+C_0)\
, \\\nonumber 
&&a_0=\left(\frac{{\cal C}_M}{\Lambda}\right)^{1/3} \!\!\!\! , 
C=\frac{\sqrt{3\Lambda}}{2} \ ,
H_0=\left(\frac{\Lambda}{3}\right)^{1/2} \!\!\!\!\!
 \coth(\sqrt{3\Lambda} t_0) \ , 
\end{eqnarray}
where $C_0=Ct_0$ and $H_0$ is the Hubble constant. With a new definition of $t$
we set $C_0=0$, and the big bang occurs at $t=0$.

 The value of the deceleration parameter \citep{MTW}
\begin{equation}
q=-\frac{\ddot a}{a}\left(\frac{\dot a}{a}\right)^{-2}
\end{equation}
decreases after the big bang, and changes sign (acceleration) at the moment of turnover
 $t=t_T$. A dot denotes the derivative with respect to the
comoving time coordinate $t$. The deceleration curve is displayed in Fig. 1.
In the limit of a vanishing cosmological constant, the turnover time $t_T$
moves to infinity, and the radius has the time dependence $a=a_0t^{2/3}$.

\begin{figure}[htb]
\begin{center}
\resizebox{10cm}{!}{\includegraphics{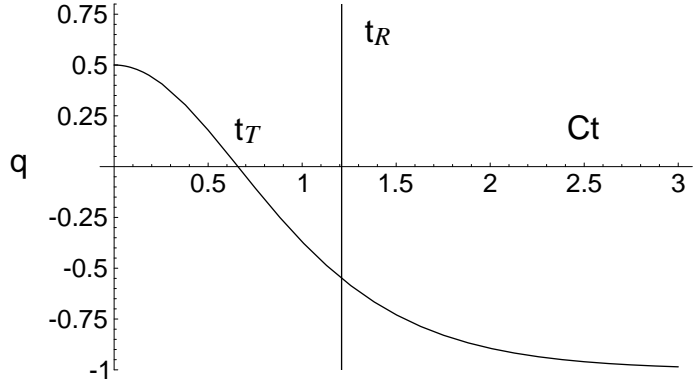}}
\end{center}
\caption{\small The deceleration parameter $q$ as a function of the dimensionless 
comoving time $Ct$. The present age of the universe is $Ct_R=1.21$ 
and the acceleration
commences at comoving time $Ct_T=0.66$ ({\it Cf.} Sec. \ref{Sec:Pi}).}
\label{f1}
\end{figure}

 The functional change of $h_{ab}$ induced by the
transformation (\ref{cootr}) is expressed by the Lie derivative of the metric $g_{ab}$
with respect to $\xi^a$, 
\begin{eqnarray}\label{transf}
&&\!\!\!\!h_{00}\rightarrow h_{00}\ , h_{0\alpha}\rightarrow h_{0\alpha} + 
\frac{b_{,\alpha}}{a}\ , h\rightarrow h +
2c^{\mu}_{\ ,\mu} + 6b\frac{\dot{a}}{a}\\
&&\!\!\!\!h_{\alpha\beta}\rightarrow h_{\alpha\beta} + c_{\alpha ,\beta} + c_{\beta,
\alpha} + 2b\frac{\dot{a}}{a}\eta_{\alpha\beta} \ .  \nonumber
\end{eqnarray}

Eq. (\ref{cons1}) is integrated
for $h^{0\alpha }$. With a coordinate transformation and using Brelot's Lemma, one 
can set \citep{White} 
\begin{equation}
h^{0\alpha }=\frac{1}{a}\nabla ^{2}C^{\alpha }\ ,
\end{equation}
where $C^{\alpha }=C^{\alpha }(x^{\beta})$ is a spatial function with 
\begin{equation}
C_{\ ,\mu }^{\mu }=0.
\end{equation}
As a consequence $h_{\ \ ,\mu }^{0\mu }=0$ holds. Gauge transformations
which preserve the above form of $h^{0\alpha }$ satisfy $\nabla^2b=0$.

Integrating Eq. (\ref{cons}) with respect to the conformal time we get 
\begin{equation}  \label{H}
h=-2\frac{\delta\rho}{\rho}+H \ ,
\end{equation}
where $H=H({x^{\alpha}})$ is an integration function.

From the perturbed field equations (\ref{Eq00}) and (\ref{Traceeq}) we have 
\begin{eqnarray}
&& h^{\prime\prime} = -2\frac{a^\prime}{a}h^\prime+ \frac{1}{2}S^{\mu\nu}_{\
\ ,\mu\nu}+\frac{1}{3}\nabla^2h \ , \\
&& \delta\rho = \frac{{a}^\prime}{a^3}{h}^\prime - \frac{1}{a^2} \left( 
\frac{1}{2}S^{\mu\nu}_{\ \ ,\mu\nu}+\frac{1}{3}\nabla^2h \right) \ .
\end{eqnarray}

From the above expressions 
\begin{eqnarray}
h^{\prime\prime} + \frac{a^\prime}{a}h^\prime + a^2\delta\rho = 0 \ .
\end{eqnarray}

Inserting here $h$ from Eq.(\ref{H}), changing to the comoving time $t$ and
using the unperturbed solution (\ref{backgr}), the equation for the density
perturbation is 
\begin{eqnarray}
\delta\ddot{\rho}+8\frac{\dot{a}}{a}\delta\dot{\rho} +3\left(\frac{\ddot{a}}{%
a}+4\frac{\dot{a}^2}{a^2} -\frac{{\cal C}_M}{6a^3} \right)\delta\rho
= 0 \ .
\end{eqnarray}

Substituting the field equations (\ref{Field0}) we have 
\begin{eqnarray}
\delta\ddot{\rho}+8\sqrt{\frac{\Lambda}{3}}\coth(Ct)
 \delta\dot{\rho} +\Lambda\left(\frac{3}{{\rm sinh}^{2}(Ct)}+5
\right)\delta\rho = 0 \ .
\end{eqnarray}

Changing to the variable 
\begin{equation}
z={\rm cosh}(2Ct)
\end{equation}
results in the homogeneous equation
\begin{eqnarray}
&&3\left( z^{2}-1\right) (z-1)\frac{d^2\delta \rho}{dz^2}\\\nonumber&&+\left(
11z^{2}-3z-8\right) \frac{d\delta \rho }{dz}+(5z+1)\delta \rho =0\ .
\end{eqnarray}

This is a special case of the Heun equation \citep{Kamke}. Application of the Kovacic
algorithm \citep{Kovacic} yields the solution 
\begin{equation}
\delta \rho =\frac{\sqrt{z+1}}{(z-1)^{3/2}}\left[ K_{1}\!-\!K_{2}\int_1^z \!\!\frac{{\rm %
d}z^{\prime}}{(z^{\prime}-1)^{1/6}(z^{\prime}+1)^{3/2}}\right] ,
\end{equation}
where $K_1=K_1(x^{\alpha})$ and $K_2=K_2(x^{\alpha})$ are space functions.
This contains the elliptic integral 
\begin{equation} \label{Iint}
I=\!\!\int_1^z \!\!\frac{{\rm d}z^{\prime}}{(z^{\prime}\!-\!1)^{1/6}(z^{\prime}\!+\!1)^{3/2}}
=\frac3{2^{2/3}}\!\!\int_0^x \!\!\left( \frac{x^{\prime}}{x^{\prime 3}\!+\!1}%
\right) ^{3/2}\!\!\!\!\!\!{\rm d}x^{\prime} \ ,
\end{equation}%
where we have introduced the new variable $x$ by 
\begin{equation}
z=2x^{3}+1.
\end{equation}%

Hence, using Eq. (\ref{H})
\begin{eqnarray}\label{h}
h=H-\coth(Ct)\left( K_{1}-K_{2}I\right) \ .
\end{eqnarray}

Taking the divergence of (\ref{Eq0al}), we obtain a total time derivative.
By Brelot's Lemma, the integral of this equation defines $\nabla^2B$ such that
\begin{eqnarray}\label{B}
\nabla^2B = \frac{1}{2}S^{\mu\nu}_{\ \ ,\mu\nu}+\frac{1}{3}\nabla^2h \ ,
\end{eqnarray}
where $B=B(x^\alpha)$ is a space function.
From (\ref{Eq00}) we have
\begin{eqnarray}
h^{\prime\prime} = -2\frac{a^\prime}{a}h^\prime+\nabla^2B\ \ .
\end{eqnarray}
Substituting here the form (\ref{h}) of $h$, the terms proportional to $K_1$ cancel
and we get
\begin{equation}
K_2=\frac{3\nabla^2B}{2^{4/3}a_0^2C^2} \ .
\end{equation}
Since $K_1$ is a $C^\infty$ function, it may be represented \citep{Friedman} by a new spatial 
function $A(x^\alpha)$ as follows, 
\begin{equation}
K_1=\nabla^2A \ .
\end{equation}

The time dependence of the density contrast $\delta\rho/\rho$ is displayed in Fig. 2. 
The contributions of the terms 
\[\delta_A=\coth(Ct)\ , \qquad
\delta_B=\coth(Ct)I
\]
proportional to the spatial functions $A$ 
and $B$, respectively, are relatively decreasing and growing. They tend to constant values 
in the asymptotic future,
\[
\lim_{t\rightarrow \infty }\delta_A%
=1 ,\quad \! \lim_{t\rightarrow \infty }\delta_B=2^{-2/3}Beta(%
{\textstyle{2 \over 3}}%
,%
{\textstyle{5 \over 6}}%
)=1.086517930 \ ,
\]
where $Beta$ is Euler's integral of the first kind.
\begin{figure}[htb]
\begin{center}
\resizebox{10cm}{!}{\includegraphics{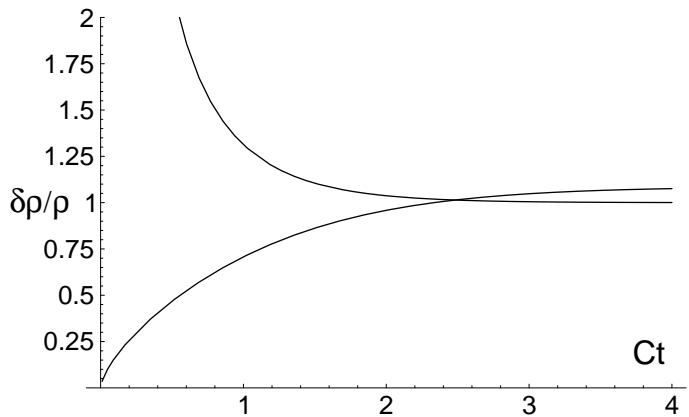}}
\end{center}
\caption{\small The relatively decreasing and growing contributions to the density contrast.}
\label{f2}
\end{figure}

These terms have the series expansion in powers of $\tau=Ct$
\begin{eqnarray}
&&\delta_A=\frac{1}{\tau }+\frac{1}{3}\tau -\frac{1}{45}\tau ^{3}+O(\tau ^{4}),\\\nonumber
&&\delta_B=2^{1/3}\frac35 \tau ^{2/3}+O(\tau ^{8/3}).
\end{eqnarray}%
In the limit of a vanishing cosmological constant, these are
the density perturbations of \citet{SW}. 

If we rewrite Eq.(\ref{Eq0al}) \ by introducing the new variable%
\[
w=\sinh (Ct)
\]%
we have 
\begin{eqnarray}\label{Sdiv}
&&\frac{\partial }{\partial w}\left( S_{\ \ ,\mu }^{\alpha \mu }-\frac{2}{3}%
h_{,}^{\ \alpha }\right) +\frac{\nabla ^{2}\nabla ^{2}C^{\alpha }}{%
a_0^{2}Cw^{4/3}\sqrt{1+w^{2}}}\\\nonumber&&-\frac{8C}{3}\frac{\nabla
^{2}C^{\alpha }}{w^{2}\sqrt{1+w^{2}}}=0\ .
\end{eqnarray}

The integration of (\ref{Sdiv}) and use of Eq. (\ref{h}) yields 
\begin{eqnarray}
&&S_{\ \ ,\mu }^{\alpha \mu }+\frac{2}{3}\coth(Ct)
\left( K_{1}-K_{2}I\right)_{,}^{\ \alpha }+\frac{8C\nabla ^{2}C^{\alpha }}{3}\frac{\sqrt{%
1+w^{2}}}{w}\nonumber \\
&&+\frac{\nabla
^{2}\nabla ^{2}C^{\alpha }}{a_0^{2}C}\int_0^w \frac{dw^{\prime}}{w^{\prime 4/3}\left(
1+w^{\prime 2}\right) ^{1/2}}+J^{\alpha }=0\ .  \label{divS}
\end{eqnarray}%
Here $J^{\alpha }=J^{\alpha }(x^\beta)$ are integration functions. 
From (\ref{transf}), the trace-free part $S_{\alpha\beta}$ has the transformation properties
\begin{eqnarray}\label{Stran}
S_{\alpha\beta}\rightarrow S_{\alpha\beta} &+& 2c_{(\alpha ,\beta)}
 - \frac{2}3\eta_{\alpha\beta}
c^{\mu}_{\ ,\mu}\ .
\end{eqnarray}
The generator $c^\alpha$ of the gauge transformations can be chosen \citep{White} 
such that the integration function in (\ref{h}) is $H=3B$
and in (\ref{divS}) we set $J^\alpha=0$. 
The normal form of the elliptic integral
\begin{equation} \label{Jint}
J=\int_0^w \frac{dw^{\prime}}{w^{\prime 4/3}\left( 1+w^{\prime 2}\right) ^{1/2}}
\end{equation}
will be given in the Appendix. The graph of $J$ is displayed in Fig. 3.
The following relation holds between the time variables:
\begin{eqnarray}
&& x=\frac{a}{a_0}=\sinh^{2/3}(Ct)=\left(\frac{z-1}{2}\right)^{1/3}=w^{2/3}\ , \\
&& x^3=\sinh^{2}(Ct)=\frac{1}{2}(\cosh(2Ct)-1)=\frac{z-1}{2}\ ,
\end{eqnarray}
where the ranges are $0<t<\infty$, $0<x<\infty$, $0<w<\infty$, $1<z<\infty$.
\begin{figure}[htb]
\begin{center}
\resizebox{10cm}{!}{\includegraphics{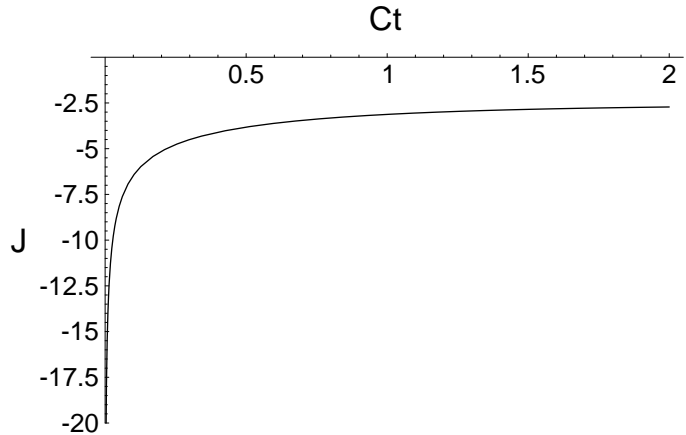}}
\end{center}
\caption{\small Time dependence of $J$}
\label{f3}
\end{figure}

  Equation (\ref{Eqalbe}) for the trace-free part of the perturbations is 
\begin{eqnarray}
&&a^{2}\ddot{S}_{\ \beta }^{\alpha }+3a\dot{a}\dot{S}_{\ \beta }^{\alpha
}-\nabla ^{2}S_{\ \beta }^{\alpha }=h_{,\ \beta }^{\ \alpha }-4B_{,\ \beta }^{\ \alpha }\label{Seq} \\\nonumber
&&-2C\nabla ^{2}(C_{\ ,\beta }^{\alpha }+C_{\beta ,}^{\ \ \alpha })
\coth(Ct)+\frac{\delta _{\ \beta }^{\alpha }%
}{3}(\nabla ^{2}h-4\nabla ^{2}B)   \\\nonumber
&&-\frac{\nabla^{2}\nabla ^{2}(C_{\ ,\beta }^{\alpha }+C_{\beta ,}^{\ \ \alpha })}{a_0^{2}C}J
\end{eqnarray}
with $h$ as given in Eq. (\ref{h}).

A particular solution for $S_{\ \beta }^{\alpha }$ can be constructed by considering the
terms with a decreasing number of Laplacians, and reads as 
\begin{eqnarray}
\stackunder{1}S{_{\ \beta }^{\alpha }} &=&\frac{\nabla ^{2}(C_{\ ,\beta }^{\alpha
}+C_{\beta ,}^{\ \ \alpha })}{a_0^{2}C}J+\frac{8C}{3}(C_{\ ,\beta
}^{\alpha }+C_{\beta ,}^{\ \ \alpha })
\coth(Ct)  \nonumber \\
&&-\left( B_{,\ \beta }^{\ \alpha }+\frac{\delta _{\ \beta }^{\alpha }}{3}%
\nabla ^{2}B\right) \frac{3I}{2^{4/3}a_0^{2}C^{2}}
\coth(Ct)\\
&&+\left( A_{,\ \beta }^{\ \alpha }+\frac{\delta _{\ \beta }^{\alpha }}{3}%
\nabla ^{2}A\right) \coth(Ct)\ .  \nonumber
\end{eqnarray}%

The general solution of Eq. (\ref{Seq}) is 
\begin{equation}
{S_{\;\beta }^{\alpha }=}{}\stackunder{0}{S}{_{\;\beta }^{\alpha }+}{}\stackunder{1}{S}{%
_{\;\beta }^{\alpha }} \ ,
\end{equation}
where ${}\stackunder{0}{S}{_{\;\beta }^{\alpha }}$ is the general solution of the
wave equation 
\begin{equation}\label{wave}
a^{2}\stackunder{0}{\ddot{S}}{_{\ \beta }^{\alpha }}+3a\dot{a}\stackunder{0}{\dot{S}}{_{\ \beta }^{\alpha
}}-\nabla ^{2}\stackunder{0}S{_{\ \beta }^{\alpha }}=0
\end{equation}
satisfying $\stackunder{0}S{_{\ \ ,\mu }^{\alpha \mu }}=0$.

Expanding the solutions of this equation in plane waves 
\begin{equation}
\stackunder{0}S{_{\ \beta }^{\alpha }}({\bf x},t)=
\int_0^\infty D_{\ \beta }^{\alpha }({\bf k},t){\rm exp}({i{\bf k\cdot x}})d^3k\ ,
\end{equation}
 the amplitude $D_{\ \beta }^{\alpha }({\bf k},t)$ of the modes satisfies
\begin{equation}\label{waveD}
a^{2}\ddot{D}_{\ \beta }^{\alpha }+3a\dot{a}\dot{D}_{\ \beta }^{\alpha
}+({\bf k\cdot k})D_{\ \beta }^{\alpha }=0
\end{equation}
and is transverse, $D^{\alpha\beta} k_\beta=0$.
Changing to the variable $x={\rm sinh}^{2/3}(Ct)$ we get the following 
equation
\begin{equation}
x(1+x^3)\frac{\partial^2D}{\partial x^2}
+\left(4x^3+\frac52\right)\frac{\partial D}{\partial x}+k^2D=0 \ ,
\end{equation}
where $k^{2}=\left(\frac{3}{2a_0C}\right)^2({\bf k\cdot k})$, with the solution
\begin{eqnarray}
D=&&x^{-3/2}\sqrt{1+4k^2x+4x^3}\\\nonumber
&&\times\left[P_k \cos(\omega_k(x))+Q_k \sin(\omega_k(x)) 
\right]  
\end{eqnarray}
and
\begin{equation}
\omega_k(x)=\sqrt{k^6+\frac{27}{64}}\int_0^x \!\!\frac{\sqrt{x^\prime}\ dx^\prime}
{\left(x^{\prime 3}+k^2x^\prime+1/4 \right)\sqrt{1+x^{\prime 3}}} \ .
\end{equation}

In summary, our solution for the linear perturbations of a $k=0 $ pressure-free 
FRW universe with a cosmological constant has the form 
\begin{equation}
p=0\ , \qquad \delta p=0\ , \qquad h^{0\alpha }=\frac{1}{a_0}{\rm sinh}^{-2/3}(Ct)\nabla ^{2}C^{\alpha }\ ,
\end{equation}

\begin{eqnarray}
h_{\alpha\beta} =&&\stackunder{0}S{_{\alpha\beta}}+A_{,\alpha\beta}
\coth(Ct)
 +{\eta_{\alpha\beta}}B\\\nonumber&&-B_{,\alpha\beta}\frac{3I}{2^{4/3}a_0^{2}C^{2}}
   \coth(Ct)
+\frac{\nabla ^{2}(C_{\alpha ,\beta }+C_{\beta ,\alpha })}{a_0^{2}C}J\\\nonumber
&& +\frac{8C}{3}(C_{\alpha ,\beta }+C_{\beta ,\alpha })
   \coth(Ct)  \ ,   \nonumber
\end{eqnarray}
\begin{equation}
\delta \rho =\frac{{\cal C}_M}{2a_0^3}
\frac{{\rm cosh}(Ct)}{{\rm sinh}^3(Ct)}
\left( \nabla^2A-\frac{3\nabla^2B}{2^{4/3}a_0^2C^2}I\right) \ ,
\end{equation}
where $A=A(x^{\alpha})$, $B=B(x^{\alpha})$, $C_{\alpha}=C_{\alpha}(x^{\alpha})$
are space functions, and the time-dependent amplitudes $I$ and $J$
are elliptic integrals,
\begin{eqnarray}
&&I=2^{-2/3}\sqrt{3\Lambda}\int^{t}_0 \frac{{\rm sinh}^{2/3}(C\tau)}{{\rm cosh}^2(C\tau)}\ {\rm d}\tau
\ , \\\nonumber
&&J=-2^{-1/3}3I(t)-3{\rm sinh}^{-1/3}(Ct){\rm cosh}^{-1}(Ct)\ .
\end{eqnarray}
The term $\stackunder{0}S{_{\alpha\beta}}$ represents gravitational waves and is a solution of the
wave equation (\ref{wave}). 
In obtaining this result we used the gauge freedom (\ref{cootr}) up to transformations with 
harmonic generator functions $b$ and $c^\mu$ satisfying 
$$\nabla^2 b = 0 \ ,\qquad \nabla^2 c^\mu = 0 \ , \qquad c^\mu_{\ ,\mu}=0 \ .$$ 
By Eq.(\ref{transf}) the form of the solution is not preserved by arbitrary gauge
transformations.

\section{The Sachs-Wolfe effect}

The Sachs-Wolfe effect is the contribution to the temperature variation $\delta T$ 
of the cosmic background radiation due to the gravitational perturbations along the
path of the photon. It can be computed \citep{SW} as follows,
\begin{equation}
\frac{\delta T}T=\frac 12\int_0^{\eta _R-\eta _E}\left( \frac{\partial
h_{\alpha \beta }}{\partial \eta }e^\alpha e^\beta -2\frac{\partial
h_{0\alpha }}{\partial \eta }e^\alpha \right)  dw \ , \label{deTT1}
\end{equation}
where $\eta _R$ and $\eta _E$ denote the time of reception and emission,
respectively, and $w$ is the affine length along the null geodesic of
propagation with tangent four-vector
\begin{equation} \label{lightprop}
\frac{dx^a}{dw}=\left( -1,e^\alpha \right) \ ,
\end{equation}
where $e^\alpha$ is the three-vector of the photon direction normalized by $e^\alpha e_\alpha =-1.$ 
We consider the contribution of the
relatively increasing mode. Then $h_{0\beta }=0$ and the second term under
the integral in Eq. (\ref{deTT1}) vanishes. The term ${}\stackunder{1}{S}{%
_{\alpha \beta }}$ has the amplitude $B_{,\alpha \beta }.$ 
By using the relation (\ref{lightprop}) in the identical decomposition 
\begin{equation}
y_{,a}\frac{dx^a}{dw}dw=y_{,\alpha }e^\alpha dw-y^{\prime }dw  
\end{equation}
for an arbitrary function $y=y(\eta,x^\alpha)$, we get dipole anisotropy contributions with 
the amplitude $B_{,\beta}e^\beta $ and gravitational redshift
terms. The time dependence of the trace part of $h_{\alpha \beta }$
is different from that in \citet{SW} thus the cancellation of the integrated terms 
does not occur here. Taken together 
the contributions to the temperature variation sum up to

\begin{eqnarray}\label{deltaTT}
&&\!\!\!\!\!\!\frac{\delta T}{T}=-\frac{3}{2^{7/3}a_0^2C^2}
 \int_0^{\eta _R-\eta _E}\!\!\! B_{,\alpha\beta}e^{\alpha}e^{\beta}
  \Delta(t) dw \\ 
 && \! = -\left[\frac{3}{2^{7/3}{a_0^2C^2}} 
 B_{,\alpha}e^{\alpha}
     \Delta(t) 
	- B\Delta_{SW}(t)
  \right]^{\eta_R}_{\eta_E} 
  \!\!\!+\left(\frac{\delta T}{T}\right)_{ ISW}
  \  \nonumber
\end{eqnarray}
at the respective events $R$ of reception and $E$ of emission,
where the transfer functions of the fluctuations are defined
\begin{equation}
\Delta(t)\!=\!a\left(\!I\coth(Ct) \right)^{_\bullet}
\!=\!a_0{\rm sinh}(Ct)^{2/3} 
    \left(I\coth(Ct) \right)^{_\bullet} \ ,
\end{equation}
\begin{equation}
\Delta_{SW}(t)=\frac{3}{2^{7/3}{a_0C^2}}{\rm sinh}(Ct)^{2/3}\dot\Delta(t)
\end{equation}
and
\begin{eqnarray}
\left(\frac{\delta T}{T}\right)_{ ISW}\!\!\!\!\!\!\!&=&\!\!
\frac{3}{2^{7/3}C^2}\!\!\int^{\eta_R}_{\eta_E}\!\!\!\!\!\!{\rm sinh}(Ct)^{2/3}\left[ 
	B{\rm sinh}(Ct)^{2/3}\dot\Delta(t)
  \right]^{_\bullet}\!\! dw \nonumber\\
&=&  -a_0C{2^{1/3}}\int^{\eta_R}_{\eta_E}B\Delta_{ISW}(t) dw 
\end{eqnarray}
is the integrated Sachs-Wolfe (ISW) term with the transfer function
\begin{equation}
\Delta_{ISW}=\frac{2\cosh^2(Ct)+3}{3\sinh^2(Ct)}I-\frac{2^{1/3}}{\sinh^{1/3}(Ct)\cosh(Ct)} \ .
\end{equation}
The terms containing $B_{,\alpha}e^{\alpha}$ have a dipole character  \citep{SW}. 
The relativistic redshift effect, represented by the second term in (\ref{deltaTT}) is displayed on
Fig. 4.
The amplitude $\Delta_{SW}$ of the temperature variations decays exponentially. In the limit of a vanishing 
cosmological constant, the amplitude has the constant value $0.1$.

\section{Physical interpretation}\label{Sec:Pi}

The combined results of the {\it WMAP} and {\it SDSS} surveys \citep{Tegmark} are for the matter 
and dark energy densities of the Universe that $\Omega_m=0.3\pm0.04$ and 
$\Omega_{\Lambda}=0.7\pm0.04$, respectively. For the Hubble parameter 
$H_0=h\cdot100km/sec/Mpc$ we use $h=0.7\pm0.04$. 
From these parameters we obtain the value of the cosmological constant 
$\Lambda=3H_0^2\Omega_{\Lambda}/c^2=0.12\cdot 10^{-51}\ m^{-2}$ and the
current age of the Universe $t_R=13.47\cdot10^9 yr$. The constant $C$ can be determined
by use of (\ref{backgr}) to be $C=2.85\cdot10^{-18} sec^{-1}$.  Taking the decoupling
to occur at redshift $z=1100$, this corresponds in the present model to 
$t_{dec}=4.66\cdot10^{5} yrs$. 

  The amplitude of the density contrast  at the decoupling is estimated in  \citet{Kolb}, to be 
$\delta\rho/\rho\approx10^{-3}$. A more recent estimate based on the $\sigma_8$ measurements  
\citep{Tegmark2} confirms this order of magnitude. The current value is determined by the amplitude 
of the relatively growing mode 
\begin{equation}
\left(\frac{\delta\rho}{\rho}\right)_{t_R}=\frac{(I\coth(Ct))_{t_R}}{(I\coth(Ct))_{t_{dec}}}
\left(\frac{\delta\rho}{\rho}\right)_{t_{dec}}\approx0.8\ .
\end{equation}
This corresponds to an amplification factor $800$. Our result is more than an order of magnitude
higher then the data in  
Table 1 of  \citet{Heath}. \citet{Lahav} find that the
growth factor depends only weakly on the value of the cosmological constant, but is much more 
sensitive to that value for higher redshifts.

  The amplitude of the temperature fluctuations $\delta T/T$ of the CMBR 
  due to the Sachs-Wolfe effect on the relatively growing mode vanishes at
$Ct=1.02$.  At the moment of reception, the amplitude is $\Delta_{SW}=-0.033$. Thus our conclusion is that the 
observed angular variation of the temperature fluctuations of the CMBR are attenuated by a factor of 3 in 
the presence of dark energy. The temporal variation 
of the SW amplitude has been observed in detailed numerical 
studies (\cite{HS}). These computations follow the variation of the multipole moments, which
obscures the effect.  

\begin{figure}[htb]
\begin{center}
\resizebox{10cm}{!}{\includegraphics{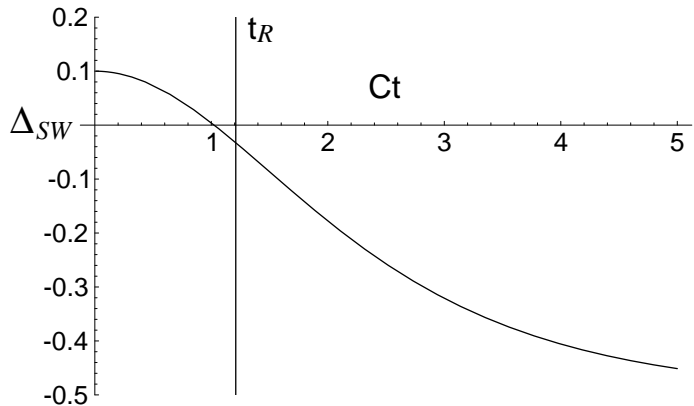}}
\end{center}
\caption{\small The dependence of the transfer function $\Delta_{SW}$ on the distance traveled by the photon from 
the event of last scattering. The decoupling can be taken to occur at $Ct_{dec}=0$ within
the error bar. Reception is indicated at $Ct_R=1.21$.}
\label{f4}
\end{figure}
The weighing function of the ISW effect is displayed in Fig. 5. The dominance of the late-time
contribution is apparent, in agreement with earlier predictions  \citep{Crittenden}. The correlation of the
ISW power spectrum with the mass distribution from galaxy counts has been verified by  \citet{Fosalba}.
\begin{figure}[htb]
\begin{center}
\resizebox{10cm}{!}{\includegraphics{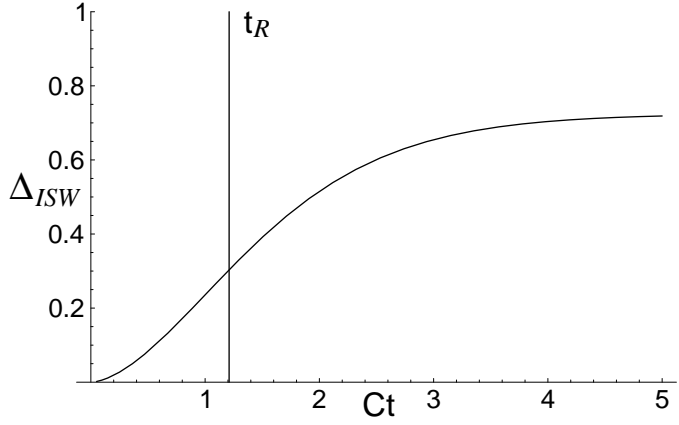}}
\end{center}
\caption{\small The weighing function of the ISW effect.}
\label{f5}
\end{figure}

\section{Concluding remarks}

  The results of our work are relevant in two major contexts: first, we find that the observed
CMBR power spectrum is attenuated by a significant overall factor due to the presence of the
cosmological constant. Our finding is consistent with the results of the  numerical studies
in \cite{HS}. This phenomenon may help understanding structure formation by taking
into account a higher level of the initial matter fluctuations. The second observation of interest
that we make is the natural complexification of the perturbed fields in the normal picture of 
Legendre. This complex description may prove useful in a treatment of quantum fluctuations,
a phenomenon that occurs also in the Fourier expansion of the wave solutions.  

\section{Appendix: Legendre normal forms}

The integral (\ref{Iint}) can be brought to the Legendre normal form \citep{Grobner} 
\begin{eqnarray}
I=&&\frac{1}{2^{1/6}}\left( 3+3^{1/2}i\right) ^{1/2}\left( E-\frac{%
3+3^{1/2}i}{6}F\right) \\\nonumber&&-\frac{2^{1/3}\left( 1-x\right) x^{1/2}}{%
(x^{3}+1)^{1/2}}  \nonumber
\end{eqnarray}%
using the mirror symmetries of the incomplete elliptic integrals 
\begin{equation}
E=E\left( \left[ 
{\textstyle{(3+3^{1/2}i)x \over 2(x+1)}}%
\right] ^{1/2}\left\vert  
{\textstyle{i-3^{1/2} \over 2}}%
\right. \right) 
\end{equation}%
and

\begin{equation}
F=F\left( \left[ 
{\textstyle{(3+3^{1/2}i)x \over 2(x+1)}}%
\right] ^{1/2}\left\vert  
{\textstyle{i-3^{1/2} \over 2}}%
\right. \right) .
\end{equation}%
To get the Legendre form of 
\begin{equation} 
J=\int_0^w \frac{dw^{\prime}}{w^{\prime 4/3}\left(1+w^{\prime 2}\right) ^{1/2}}=\frac{3}{2}\int_0^x \frac{%
dx^{\prime}}{x^{\prime 3/2}\left( 1+x^{\prime 3}\right) ^{1/2}}
\end{equation}
we use the variable $w=x^{3/2}$ in the integral. The normal form is \citep{Grobner} 
\begin{eqnarray}
J=&-&\frac{3}{2}2^{1/2}(3+3^{1/2}i)^{1/2}\left( E-\frac{%
3+3^{1/2}i}{6}F\right)\\\nonumber &-&3\frac{(x^{2}-x+1)^{1/2}}{(x+1)^{1/2}x^{1/2}}\ .
\end{eqnarray}
The following relation holds between the incomplete elliptic integrals:
\begin{eqnarray}
J+\frac{3}{2^{1/3}}I=-\frac{3}{x^{1/2}(1+x^3)^{1/2}}\ . 
\end{eqnarray}

\section{Acknowledgements}
We thank professor J\"urgen Ehlers for suggesting improvements to the
manuscript.
This work was supported by OTKA no. T031724 and TS044665 grants.

\end{document}